\documentclass[journal,transmag]{IEEEtran}

\ifCLASSINFOpdf
\else
\fi

\usepackage{graphicx}
\usepackage{enumitem}
\usepackage{cite}
\usepackage{amsmath}
\usepackage{dsfont}
\usepackage{siunitx}

\usepackage{tikz}
\usepackage{pgfplots}
\pgfplotsset{width=7cm,compat=newest}
\usetikzlibrary{colorbrewer,patterns}
\usepgfplotslibrary{colorbrewer}
\usetikzlibrary{external}
\pgfplotsset{cycle list/Set1}

\pgfplotsset{compat=1.13}

\begin{document}
\title{GPU Accelerated Atomistic Energy Barrier Calculations of Skyrmion Annihilations}

\author{\IEEEauthorblockN{Paul Heistracher\IEEEauthorrefmark{1},
Claas Abert\IEEEauthorrefmark{1},
Florian Bruckner\IEEEauthorrefmark{1}, 
Christoph Vogler\IEEEauthorrefmark{1}
and Dieter Suess\IEEEauthorrefmark{1}
}
\IEEEauthorblockA{\IEEEauthorrefmark{1} Christian Doppler Laboratory, Faculty of Physics, 
	University of Vienna, Austria}
\thanks{Corresponding author: 
	
	P. Heistracher (email: paul.thomas.heistracher@univie.ac.at).}}

\IEEEtitleabstractindextext{%
\begin{abstract}
We present GPU accelerated simulations to calculate the annihilation energy of magnetic skyrmions in an atomistic spin model considering dipole-dipole, exchange, uniaxial-anisotropy and Dzyaloshinskii-Moriya interactions using the simplified string method. The skyrmion annihilation energy is directly related to its thermal stability and is a key measure for the applicability of magnetic skyrmions to storage and logic devices.  We investigate annihilations mediated by Bloch points as well as annihilations via boundaries for various interaction energies.
Both processes show similar behaviour, with boundary annihilations resulting in slightly smaller energy barriers than Bloch point annihilations.
\end{abstract}

}

\maketitle

\IEEEdisplaynontitleabstractindextext

\IEEEpeerreviewmaketitle

\section{Introduction}

Isolated magnetic skyrmions have recently attracted a lot of attention in the scientific community as a
potential candidate for storage and for logic devices \cite{Fert2013,Sampaio2013,Romming636,ThermalStability}. It has been demonstrated that an isolated skyrmion can be a
stable configuration in a nanostructure, that it can be locally nucleated by injection of a spin-polarized current and that
it can be displaced by current-induced spin torques, even in the presence of large defects \cite{Sampaio2013}. According experiments
demonstrate the applicability of topological charge as a carrier of information by showing the feasibility to write and
erase such spin textures in a controlled fashion using local spin-polarized currents from a scanning tunnel microscope
\cite{Romming636}. Single skyrmions or chains of skyrmions can be nucleated as a metastable state in thin films and the controlled
creation and annihilation of these isolated skyrmions is opening a path to new concepts of spintronic devices.

In these applications, the temporal stability of the encoded information is an important measure for their feasibility. The
energy required to create or annihilate skyrmions is directly related to their stability and is therefore of high interest.
Due to their topological protection, skyrmionic spin configurations cannot be continuously deformed to other magnetic
configurations such as spin spirals or ferromagnetic states. This property gives rise to their comparably high
annihilation energy. 

Even though micromagnetic simulations are capable of describing skyrmion configurations, they cannot account for
skyrmion creation or annihilation processes mediated by a change of topological charge via Bloch points.
Such processes would be in contradiction with the main assumption of micromagnetics, which states that the magnetic field can be
approximated as a continuous field. As a consequence, we consider an atomistic spin model to account for such skyrmion
creation and annihilation processes. The atomistic model implies a considerably higher computational effort due to the
increased number of elements in order to account for the high spacial resolution required by the atomic lattice.

For this reason we develop a GPU accelerated atomistic code which solves the Landau-Lifshitz-Gilbert
equation for an atomistic spin model considering dipole-dipole, exchange, uniaxial-anisotropy and Dzyaloshinskii-
Moriya (DM) interactions. We further apply the \textit{improved string method} \cite{StringMethod} in order to calculate the creation and
annihilation energy of magnetic skyrmions.
For an initial transition path between two selected magnetic configurations representing local minima in the energy landscape, the method allows calculation of the most-probable path accessible from the initial string.
This path corresponds to the minimum energy path (MEP) and provides the energy barrier between the two magnetic field configurations. 
For the calculation of skyrmion creation and annihilation energies we choose a meta-stable skyrmionic state as the first configuration and a ferromagnetic state the latter configuration.

The performance gain achieved by using a GPU accelerator enables a comprehensive parameter study for various values of anisotropy and DM interaction energies, in which we compare two different annihilation processes. 
In the first case, the initial skyrmion collapses via a Bloch point. In the second case, the skyrmion is annihilated over the boundary of the system.
Both processes show similar behaviour, whereby the boundary annihilation leads to slightly smaller energy barriers. 

\section{Atomistic Spin Model}\label{sec:asm}
We describe the magnetic properties at an atomic level considering classical spins $\boldsymbol{S}_i$ on a simple cubic lattice subject to interactions given by the Hamiltonian
\begin{multline}\label{eq:hamiltonian_explicit}
\mathcal{H} = 
 \sum_{\langle i,j\rangle} \big[- J \boldsymbol{\hat{s}}_{i} \cdot \boldsymbol{\hat{s}}_{j} + d (\boldsymbol{\hat{r}}_{ij} \times \boldsymbol{\hat{z}}) \cdot ( \boldsymbol{\hat{s}}_{i} \times \boldsymbol{\hat{s}}_{j})\big]
- k \sum_{i} (\boldsymbol{\hat{s}}_{i} \cdot \boldsymbol{\hat{z}})^2 
\\ - \frac{\mu_0}{8 \pi} \sum_{i,j\neq i}^{} \frac{3(\boldsymbol{S}_i \cdot \boldsymbol{\hat{r}}_{ij})(\boldsymbol{S}_j \cdot \boldsymbol{\hat{r}}_{ij}) - \boldsymbol{S}_i \cdot \boldsymbol{S}_j }{r^3_{ij}},
\end{multline}
where $\boldsymbol{\hat{s}}_{i}=\boldsymbol{S}_i/\|\boldsymbol{S}_i\|$, as in \cite{rohart_path_2016, PhysRevB.94.174418,ThermalStability}. The first two terms describe the Heisenberg interaction with exchange energy $J$ and the Dzyaloshinskii-Moriya interaction with interaction constant $d$, respectively. The summation is performed over nearest neighbours $\langle i,j\rangle$ and $\boldsymbol{\hat{r}}_{ij}$ denotes the unit vector pointing from site $i$ to site $j$.
The third term accounts for the uniaxial anisotropy with interaction constant $k$ and the last term is the dipolar coupling, where $r_{ij}$ is the distance between sites $i$ and $j$.
For the calculations performed in this work, we restrict both the anisotropy and the DM to the out-of-plane unit vector $\boldsymbol{\hat{z}}$, as implied in the Hamiltonian. 
The samples are limited by free boundary conditions and we consider a simple cubic lattice, resulting in a data structure that is well suited for GPU computing.

\subsection{Effective Field}
For calculations that describe the dynamics of the system, we are interested in computing the effective field resulting from the Hamiltonian. 
Denoted as $\boldsymbol{H}^i_{\text{eff}}$, the effective field at atomic site $i$ is calculated by the negative partial derivative of the Hamiltonian with respect to the spin $\boldsymbol{S}_i$: 
\begin{equation}\label{eq:Heff}
	\boldsymbol{H}^i_{\text{eff}} 
	= - \frac{1}{\mu_0} \frac{\partial \mathcal{H}}{\partial \boldsymbol{S}_i}
\end{equation}

It is common to split up the effective field into several parts, each of which describes one respective interaction:
	$\boldsymbol{H}^i_{\text{eff}}=\boldsymbol{H}^i_{\text{exc}}+\boldsymbol{H}^i_{\text{DM}}+\boldsymbol{H}^i_{\text{ani}}+\boldsymbol{H}^i_{\text{dip}}$.
The contribution to the effective field originating from the exchange interaction $\boldsymbol{H}^i_{\text{exc}}$ is given by
\begin{equation}
	\boldsymbol{H}^i_{\text{exc}} 
	= \frac{J}{\mu_0 \|\boldsymbol{S}_i\|} \sum_{j} \boldsymbol{\hat{s}}_{j},
\end{equation}
where the summation considers nearest neighbours of site $i$. 
The effective field resulting from the DM interaction $\boldsymbol{H}^i_{\text{DM}}$ is computed likewise by applying equation (\ref{eq:Heff}) and yields
\begin{equation}
	\boldsymbol{H}^i_{\text{DM}} = - \frac{d}{\mu_0  \| \boldsymbol{S}_i \|} \begin{pmatrix} s_z^{i+x} - s_z^{i-x} \\ s_z^{i+y} - s_z^{i-y} \\ - s_x^{i+x} + s_x^{i-x} -s_y^{i+y} + s_y^{i-y} \end{pmatrix},
\end{equation}
where $s_x$, $s_y$ and $s_z$ denote the $x$, $y$ and $z$-component of the spin vector $\boldsymbol{\hat{s}}$ 
and the superscripts ${i+x}$, ${i-x}$, ${i+y}$ and ${i-y}$ refer to the directly neighbouring spins in positive and negative $x$- and $y$-direction, respectively. 

 The uniaxial anisotropy interaction contributes the field $\boldsymbol{H}^i_{\text{ani}} $ and reads
\begin{equation}\label{eq:Hani}
\boldsymbol{H}^i_{\text{ani}} 
= \frac{2 k}{\mu_0 \|\boldsymbol{S}_i\|} (\boldsymbol{\hat{s}}_{i} \cdot \boldsymbol{\hat{z}}) \boldsymbol{\hat{z}}.
\end{equation}

The dipole-dipole interaction is the only global interaction in the Hamiltonian of equation (\ref{eq:hamiltonian_explicit}) and by far the most computationally demanding.
With regard to a more efficient calculation, we can express the dipole-dipole field $\boldsymbol{H}^i_{\text{dip}}$ as a discrete convolution: 
\begin{equation}\label{eq:dipol-dipol-field}
\boldsymbol{H}^i_{\text{dip}} = \|\boldsymbol{S}_i\| \sum_{j} \boldsymbol{\tilde{D}}_{ij} \boldsymbol{\hat{s}}_j
\end{equation}
Here $\boldsymbol{\tilde{D}}_{ij}$ denotes the dipole-dipole tensor for atomic sites $i$ and $j$ given by

\begin{equation}\label{eq:Ddip}
\boldsymbol{\tilde{D}}_{ij} = 
\frac{1}{4 \pi}
\begin{pmatrix}
\frac{3 r_x^2}{r^5} - \frac{1}{r^3 } & \frac{3 r_x r_y}{r^5} & \frac{3 r_x r_z}{r^5}\\ 
\frac{3 r_x r_y}{r^5} & \frac{3 r_y^2}{r^5} - \frac{1}{r^3 } & \frac{3 r_y r_z}{r^5}\\ 
\frac{3 r_x r_z}{r^5} & \frac{3 r_y r_z}{r^5} & \frac{3 r_z^2}{r^5} - \frac{1}{r^3 }
\end{pmatrix}
,
\end{equation}
where $r_x$, $r_y$, $r_z$ are the $x$-, $y$-, $z$-components of the vector $\boldsymbol{r}_{ij}$, respectively, and $r$ its norm.
We emphasize the convolutional structure of the dipole-dipole field in equation (\ref{eq:dipol-dipol-field}). 
Sorting the entries in a circular manner allows a computation using fast Fourier transforms \cite{BertramFFT}.
Exploiting this property drastically reduces the computational complexity from $\mathcal{O} (N^2)$ to $\mathcal{O} (N \log N)$, where $N$ is the number of atomic sites.

After calculating the effective field, we directly obtain the energy value of the Hamiltonian $E$ by evaluating
\begin{equation}
	E =-\frac{\mu_0}{2} \sum_i \boldsymbol{H}^i_{\text{eff}} \cdot \boldsymbol{S}_i.
\end{equation}

\subsection{Spin Dynamics}
The central equation describing the dynamics of the system is the Landau-Lifshitz-Gilbert (LLG) equation
\begin{equation}\label{eq:LLGatom}
\frac{\partial \boldsymbol{\hat{s}}}{\partial t}=- \frac{\gamma}{1+\alpha^2} \boldsymbol{\hat{s}} \times \boldsymbol{H}_{\text{eff}} - \frac{\alpha \gamma}{1+\alpha^2} \boldsymbol{\hat{s}} \times (\boldsymbol{\hat{s}} \times \boldsymbol{H}_{\text{eff}}),
\end{equation}
where $\gamma= \SI{2.2127615e5}{m /A s} $ is the reduced gyromagnetic ratio and $\alpha$ a phenomenological damping constant. 
The first term in the LLG equation describes the precession of the magnetization and originates from the quantum mechanical interaction of the atomic spin with the effective field. 
The second term is introduced due to the relaxation of the magnetization and accounts for energy dissipation which causes the spins to align along the field direction with a characteristic coupling strength given by $\alpha$.
This model is capable of describing the dynamics of skyrmionic spin configurations at an atomistic scale.

\section{String Method}\label{sec:string}
In order to calculate the annihilation energy of magnetic skyrmions, we apply the \textit{simplified and improved string method} \cite{StringMethod}. 
This method allows the calculation of the most probable transition path between local minima of the energy landscape. These paths correspond to minimum energy paths (MEPs), which are paths in configuration space where the potential force is parallel to the path at every point. 
The calculation of an MEP allows the identification of the relevant saddle point, which is used to determine the energy barrier for the respective transition.

In our case, such a string is discretized by 
a number of magnetic field configurations $\{\varphi_k(t), k=0,1,\cdots,K-1 \}$, in the following referred to as \textit{images}. 
These images are intermediate replicas of the system used to create a discrete representation of the transition path. 
As the first image, a skyrmionic field configuration is selected and all energies are expressed in relation to this image. 
As the last image we select a ferromagnetic state. 
For a given set of initial images, the string is evolved iteratively by the following two-step procedure:

In the first step, each of the images on the string is evolved over a given time interval $\Delta t$ according to 
\begin{equation}\label{eq:varphi}
\dot{\varphi_k}=-\nabla V (\varphi_k),
\end{equation}
where $\dot{\varphi_k}$ is the time derivative of $\varphi_k$ and $-\nabla V (\varphi_k)$ is the potential force acting on the system. In our case, the potential force is given by the energy-dissipation term of the LLG equation, which is the second term of equation (\ref{eq:LLGatom}). 
A short derivation justifying this assumption is given in the appendix.
For time integration we use explicit Runge-Kutta methods with adaptive step-size control \cite{Numerical}, such as the fourth-order Runge-Kutta-Fehlberg method or the fourth-order Dormand-Prince method.

In the second step, 
the images are redistributed along the string by a re-parametrization procedure.
The standard choice is enforcing parametrisation by equal arc length. 
After the first step, we obtain images $\{\varphi_k^*\}$ with a non-uniform spacing $\{ \alpha_k^*\}$ and now these values are interpolated onto a uniform mesh. 
The arc length $s_k$ of the current images $\{\varphi_k^*\}$ is calculated by 
\begin{equation}\label{eq:arclength}
	s_0=0, \ s_k=s_{k-1}+ \|\varphi_k^*-\varphi_{k-1}^*\|, \ k=1,2,\cdots,K-1.
\end{equation} 
The non-uniform spacing $\{ \alpha_k^*\}$ is obtained by normalization of $\{ s_k\}$:
\begin{equation}
	\alpha_k^*= \frac{s_k}{s_N}
\end{equation}
Interpolation of $\{\varphi_k^*\}$ onto the uniform grid points $\alpha_k=k/K$ yields the parametrized images $\{\varphi_k\}$.

\begin{figure}[h!]
	\includegraphics{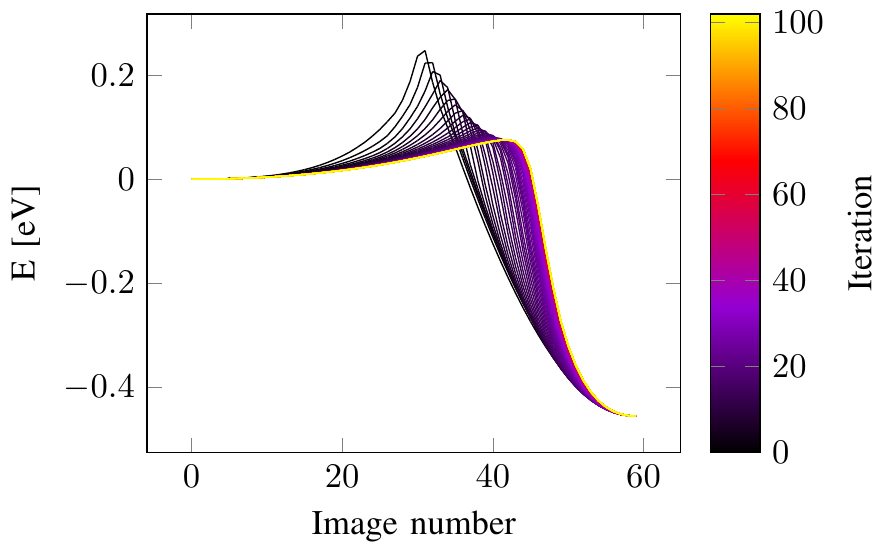}
	\caption{Energy curves of a typical annihilation process given for each iteration of the string method. The first image is the initial skyrmionic field configuration and the last image a ferromagnetic state. The energy of each image is given with respect to the first image.}
	\label{fig:E_curves}
\end{figure}

These two steps are repeated until convergence of the string is reached and the energy barrier is obtained by the energy of the energetically highest image. 
Fig. \ref{fig:E_curves} shows the energy curves obtained in each iterative step of the string method for a typical Bloch point annihilation process.
In many cases, a further refinement of the saddle-point by techniques such as the \textit{climbing image} method is achieved. 
Given the smooth plateau of the energy curve and the high number of images used in our calculations, further refinements of the saddle-point are omitted.

\section{Energy Barrier Calculations}
In the following, we present skyrmion annihilation calculations performed with our self-developed GPU accelerated simulation software, which is implemented with extensive use of the \textit{ArrayFire accelerated computing library} \cite{Yalamanchili2015}.

We model a mono-layer of $112 \times 112 \times 1$ spins in the $xy$-plane on a simple cubic lattice with lattice constant $a=\SI{0.2715}{nm}$, subject to the interactions described by the Hamiltonian of the atomistic spin model introduced in section \ref{sec:asm}. 
The spins are assumed to have a magnetic moment of $\|S_i\|=\SI{2.3737 }{\mu_B}$, where $\mu_B$ is the Bohr magneton, and the exchange energy is set to $J=\SI{54.23}{meV}$. 
The values of the DM interaction energy $d$ and the uniaxial-anisotropy energy $k$ are varied to obtain a phasediagram. We discretize each string with 60 images and evolve each image for $\Delta t=\SI{5e-14}{s}$ in each iteration step. 
Given a typical number of iterations between $100$ and $10000$, one string calculation is comparable to evolving a single configuration for $\SI{3e-10}{s}$ to $\SI{3e-8}{s}$.

\begin{figure}[h!]
	\includegraphics{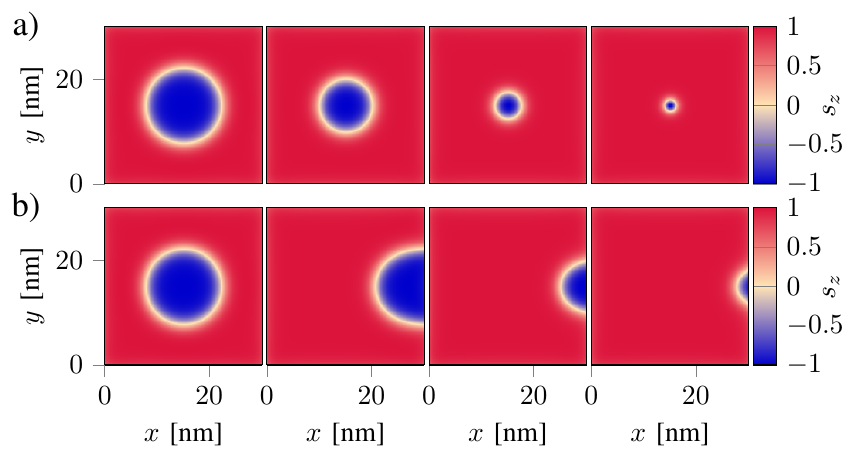}
	\caption{Selected images of skyrmion annihilation processes as obtained by the string method. a) Bloch point mediated skyrmion annihilation. b) Boundary annihilation process. Colourized according to the magnitude of the out of plane spin direction.}
	\label{fig:MEPs}
\end{figure}

Two possible MEPs between the initial radial symmetric N\'eel skyrmion configuration and the ferromagnetic state are considered. 
In the first case, the skyrmion annihilation process is mediated by a Bloch point. In the converged string, the skyrmion shrinks in size for each image until a Bloch point configuration is obtained that collapses into the ferromagnetic state. 
Fig. \ref{fig:MEPs} a) depicts four selected images of such a transition path. 
In the second case, the skyrmion is annihilated via a boundary of the system. For each image of the converged string, the skyrmion is shifted towards the boundary and finally disappears. Four selected images of a typical boundary annihilation process are shown in Fig. \ref{fig:MEPs} b).

\newcommand\SkyrmTikScale{.8}
\newcommand\SkyrmWidth{2}
\newcommand\SkyrmSpacingfactor{1.03}
\newcommand\Skyrmrowb{1.15}

The choice of the initial transition path is a delicate task. 
If multiple MEPs are present between the first and the last image of the transition, the optimization will most likely result in an MEP which is directly accessible from the initial transition path without increasing the energy. Accordingly, the MEP obtained by the converged string is the one closest to the initial transition path. 
In the Bloch point annihilation process, the images of the initial string are obtained by interpolation between the skyrmion configuration and the ferromagnetic state. 
This causes the skyrmion to shrink in each successive image, resulting in a rapidly converging string. 
For the boundary annihilation processes, the initial string is obtained by shifting the skyrmion in the direction of any boundary for each successive image. 
Initial annihilation paths over a corner are energetically higher and evolve into paths with an annihilation over the edges, where all four edges result in the same energy barrier.

\begin{figure}[h!]
	\includegraphics{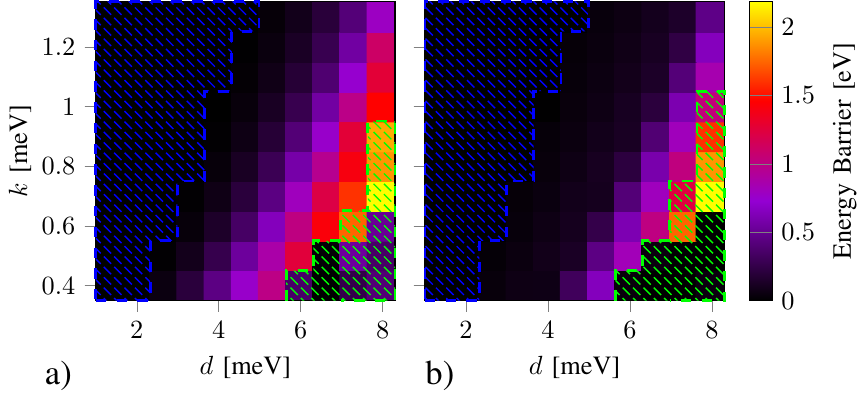}
	\caption{Phasediagrams of skyrmion annihilation energies as a function of the DM interaction energy $d$ and uniaxial-anisotropy
		energy $k$.
		a) Bloch point-mediated annihilation processes.
		b) Skyrmion annihilations mediated by boundaries.
		For values of $d$ and $k$ marked by the dashed blue line, no stable skyrmion configuration is obtained and the energy barrier is zero.
		For values within the dashed green line, the ferromagnetic configuration is no longer a local energetic minimum and evolves to a skyrmionic configuration itself, resulting in higher order transitions.}
	\label{fig:phasespace_combined}
\end{figure}

For each of these two annihilation processes, we perform calculations for various values of the DM interaction energy $d$ and the
uniaxial-anisotropy energy $k$ and obtain two phasediagrams shown in Fig. \ref{fig:phasespace_combined}.
We encounter two opposing trends, where high values of $d$ favour the formation of domain walls, while high values of $k$ suppress domain walls. 
This is due to the fact that the uniaxial-anisotropy interaction favours spins directed parallel or antiparallel to the distinguished axis whereas the DM interaction causes the spins to tilt with respect to each other.
In the region with low values of $d$ and high values of $k$ surrounded by a dashed blue line, the creation of skyrmions is suppressed and no stable skyrmion configuration is obtained. In the region with high values of $d$ and low values of $k$ marked by a dashed green line, the
ferromagnetic configuration used as the last image of the string is no longer a local energetic minimum and evolves to a
skyrmionic configuration itself. Therefore the energy barriers in these regions do not represent skyrmion annihilation processes but higher order transitions from one skyrmion configuration into another.
Skyrmion annihilation processes mediated by either
Bloch points or boundaries are observed in the intermediate region and feature straight lines of equal energies in the phase diagram indicating a linear relation between the two trends.
Boundary annihilations result in slightly lower energy barriers than Bloch point annihilations.

\begin{figure}[h!]
	\includegraphics{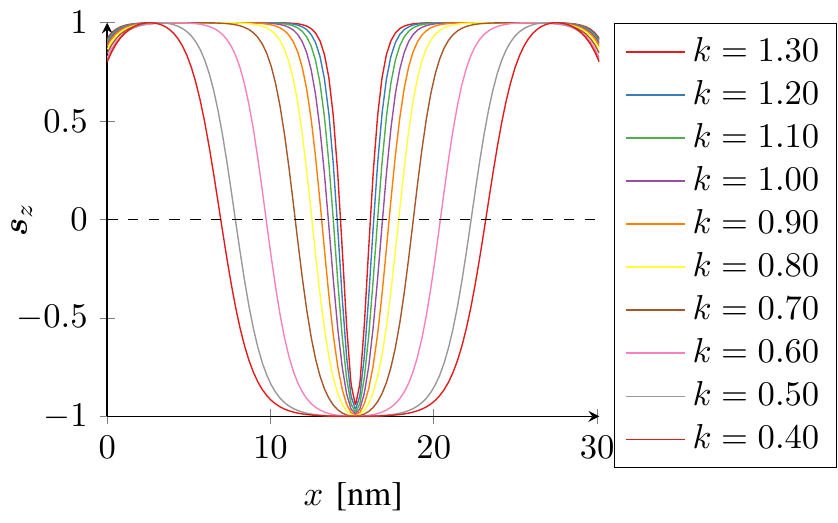}
	\caption{Skyrmion profile as a function of the anisotropy energy $k$ (in \SI{}{meV}). Cross-section through the skyrmion center parallel to the $y$-axis with a DM interaction energy of $d=\SI{5.30}{meV}$.}
	\label{fig:cross-section}
\end{figure}

\newcommand\PhasespaceWidth{4 cm}
\newcommand\PhasespaceHeigth{4}
The size of the initial skyrmion used in the string method is proportional to the uniaxial-anisotropy energy $k$. For lower values of $k$, the skyrmion diameter increases, as shown in Fig. \ref{fig:cross-section}. The skyrmion diameter is identified by the distance between the intersections of the $z$-component of the spin $s_z$ with the dashed line at zero.

\appendix
\section{}\label{sec:appx_a}
In section \ref{sec:string}, we assume the energy-dissipation term of the LLG equation 
as the potential force acting on the images of the string. 
In general, the potential force $\boldsymbol{F}=- \nabla V$ is the negative gradient of the potential energy $\mathcal{H}$ and points in the direction of the steepest decent on the energy landscape.
As we require conservation of the magnetization norm,  the potential force $\boldsymbol{F}$ is restricted to the component perpendicular to the magnetization. 
Without this constraint, the energetic minimum obtained by following the full negative gradient would yield magnetic moments of length zero. For the potential force acting on each image we therefore write

\begin{equation}\label{eq:force}
\begin{split}
\boldsymbol{F}(\boldsymbol{\hat{s}}_i)&=-\Big(\frac{\partial \mathcal{H}}{\partial \boldsymbol{\hat{s}}_i}\Big)_{\perp}\\
& =-  \Bigg[\frac{\partial \mathcal{H}}{\partial \boldsymbol{\hat{s}}_i}-\Big(\frac{\partial \mathcal{H}}{\partial \boldsymbol{\hat{s}}_i}\Big)_{\parallel} \Bigg] \\
&= -\Big(\boldsymbol{\hat{s}}_i \cdot \boldsymbol{\hat{s}}_i\Big) \frac{\partial \mathcal{H}}{\partial \boldsymbol{\hat{s}}_i} + 	\Big(\boldsymbol{\hat{s}}_i \cdot \frac{\partial \mathcal{H}}{\partial \boldsymbol{\hat{s}}_i}\Big) \cdot \boldsymbol{\hat{s}}_i\\
&=\boldsymbol{\hat{s}}_i \times \Big(\boldsymbol{\hat{s}}_i \times \frac{\partial \mathcal{H}}{\partial \boldsymbol{\hat{s}}_i} \Big) \\
&=- \boldsymbol{\hat{s}}_i \times \Big(\boldsymbol{\hat{s}}_i \times \mu_0 \|S_i\| \boldsymbol{H}_{\text{eff}} \Big),
\end{split}
\end{equation}
where we used the identity $a \times (b \times c) = (a \cdot c) \cdot b - (a \cdot b) \cdot c$ and identified the effective field $\boldsymbol{H}^i_{\text{eff}}$ according to equation (\ref{eq:Heff}).
The final expression in equation (\ref{eq:force}) differs from the second term of the LLG in equation (\ref{eq:LLGatom}) only by a constant factor which changes the arbitrary time scale and we adopt the prefactors of the energy-dissipation term of the LLG. 

\section{}

\section*{Acknowledgment}

The  financial  support  by  the  Austrian  Federal  Ministry
for  Digital  and  Economic  Affairs  and  the  National  Foundation for Research, Technology and Development is gratefully acknowledged.

\ifCLASSOPTIONcaptionsoff
  \newpage
\fi

\bibliographystyle{IEEEtran}
\bibliography{ref}

\end{document}